\documentclass[twocolumn]{aa}
\usepackage{graphicx}

\begin{document}

\title{Can stellar wobble in triple systems mimic a planet?}

\subtitle{}

\author{ J. Schneider
  \thanks{\email{jean.schneider@obspm.fr}}
  \and J. Cabrera
}

\offprints{J. Schneider}

\institute{Observatoire de Paris-Meudon, 92195, Meudon Cedex, France}

\date{Received ; accepted }

\abstract{The first extrasolar planets have been detected by the 
measurement of the wobble of the parent star. This wobble leads to the 
periodic modulation of three observables: the radial velocity, the position 
on the sky and the time of arrival of periodic signals. We show that the same 
wobble, and therefore the same modulation of the three observables, can be due
 to the presence of a more distant binary stellar companion. Thus, the 
observation of the wobble does not, by itself, constitute a proof of a planet 
detection. In particular, astrometric confirmation of a wobble does not 
necessarily provide a sufficient proof of the existence of a planet candidate 
detected by radial velocity. Additional conditions, which we discuss here, 
must be fulfilled. We investigate the observed wobble for the  planet 
candidates already detected and we find that, for each case, 
a wobble due to a binary 
stellar companion can be excluded.

But for apparent Saturn-like planets in wide orbits, there may be an 
ambiguity in 
future detections, especially in spaceborne astrometric missions.
We conclude that, in some cases, a definitive proof for the presence of a 
planet requires further observations such as direct imaging. 

\keywords{stars: planetary systems -- astrometry -- celestial mechanics}
}

\titlerunning{Can stellar wobble in triple systems mimic a planet?}
\authorrunning{Schneider \& Cabrera}

\maketitle

%
\section{Introduction}

The detection of the first extrasolar planets rests on an indirect method, 
namely the measurement of the reflex motion of the parent star. In  cases 
where only the wobble is detected, one can ask
whether the detection of radial velocity (RV) variations 1/ are indeed due to 
a stellar wobble and not to other effects (such as stellar rotation
or variable stellar activity) 2/ are indeed due to a planet and not to other 
dynamical effects 3/ that the companion is indeed a planet (and not for
instance a planetary mass black hole or strange matter object). In a few cases,
the planet detection is confirmed by (or was preceded by) the detection of 
a transit of a planet, but the question of the planetary explanation of the
wobble remains a priori open for the other candidates.

Here we consider the case where the wobble is real but due to the perturbation
by a distant binary star first suggested by Schneider (\cite{schneider99}).

\section{An approximation: restricted 3 body problem for hierachical systems}

\begin{figure}[b]
  \begin{center}
  \includegraphics[width=0.8\linewidth]{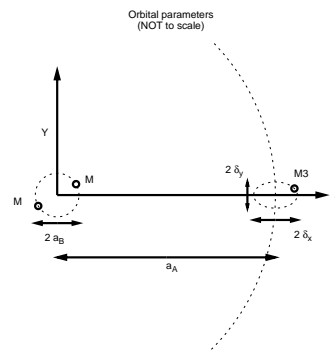}
  \end{center}
  \caption{Orbital elements of the system (not to scale)}
  \label{fig:esquema}
\end{figure}

Consider a triple hierachical system consisting of a binary system (with
masses $M_{1}  = M_{2} = M;$ separation between components $2\,a_{B}$)
and a third companion orbiting the center of mass of the system in a
bigger circular orbit (mass $M_{3};$ radius of the orbit
$a_{A}$). The perturbation caused by the binary system on the orbit of
the third companion can imitate the perturbation caused by a planet
around the latter star. In the appendix we have derived the equations
of motion of such an approximation. The result of the perturbation of
the binary system is an elliptical pe\-rio\-dic motion superimposed to the
bigger orbit of the third star. Studying this perturbation in the
plane of the orbit of the third star, let the X axis lay from the
center of mass of the binary system to the target star, and the Y axis
perpendicular to it (see fig \ref{fig:esquema}), the magnitude
of this perturbation  (see appendix) in both axis is:

\begin{equation} \label{eq:wobble_noappendix}
  \delta_{x} = 4.5 \,  \frac{a_{B}^{5}}{a_{A}^{4}}
\qquad \qquad
  \delta_{y} = 3  \, \frac{a_{B}^{5}}{a_{A}^{4}}
\end{equation}

Our approximation takes as starting points the sizes of the orbits of
the binary system ($a_{B}$) and the third star ($a_{A}$) together with
the mass of the stars in the binary system ($M$) and gives the
amplitude of the perturbation ($\delta_{x}; \, \delta_{y}$) which the
binary system causes in the movement of the third star. 

Stellar wobble and planetary companion produce, in principle, the same
perturbation: a periodical elliptical motion which can be measured
either by radial velocity or astrometry.

We have made simulations of such triple systems with the code
\emph{KAPPA}. Taking as initial parameters $a_{B}$, $a_{A}$ and $M$
(see figure \ref{fig:esquema}) we can calculate the initial position
and velocities for the three bodies which are required by the
code (using \ref{eq:def_rMi}, \ref{eq:pos_M3} and
\ref{wobble}). Finally, we compare the results obtained in the
simulation with those expected according to \ref{eq:solution_systeme}.

\begin{figure}[ht]
  \begin{center}
  \includegraphics[width=0.8\linewidth]{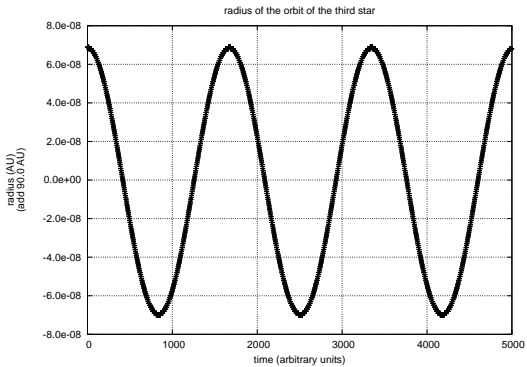}
  \end{center}
  \caption{Variation of the distance from the third star to the center of 
mass of the binary system.}  \label{fig:radius}
\end{figure}

As an example, we are going to choose $a_{B} = 1 \, \mathrm{AU}; \,
a_{A} = 90 \, \mathrm{AU}; \, M = 1 \, M_{\sun}$. In
fig.\ref{fig:radius} we see the radius of the orbit of the third star
around the center of mass of the system. In a period of three years,
it oscillates three times (half the period of the binary system, as
expected). The radius fits very well to \ref{eq:solution_systeme}.

\section{Imitating a planet}

The perturbation of the binary system can induce a motion in the
target star imitating a planet. In other words, we see a star with a
periodic elliptical wobble. We suppose that this wobble is caused by
the orbit of the star around the center of mass of the system
star-planet. However, this wobble can be in fact caused by a far
binary system to which our target star is gravitationally linked. In
our model, this motion will have the amplitude given by
\ref{eq:wobble_noappendix} and the following period:

\begin{equation} \label{eq:period}
P_{*} = P_{\mathrm{pl}} = \frac{1}{2} \, P_{\mathrm{binary}} = 
  \frac{2 \pi a_{B}^{3/2}}{\sqrt{G M}}
\end{equation}

where $M$ is the mass of each star of the binary system, $a_{B}$ half
the semi-major axis of the orbit of the binary system, as described
previously. The subscripts $*$ and \emph{pl} refer to the target star
and the supposed planet respectively. Other parameters are:

\begin{eqnarray} \label{eq:sem_maj_ax_pl}
& a_{\mathrm{pl}} & = \left( \frac{G M_{*}}{4 \pi^{2}} \, 
  P_{\mathrm{*}}^{2} \right) ^{1/3}
\\ \label{eq:eccentricity}
& e & = \sqrt{ 1 - \left( \frac{\delta_{y}}{\delta_{x}} \right)^{2}  }
\\ \label{eq:mass}
& M_{\mathrm{pl}} & =  M_{*} \, \frac{\delta_{x}}{a_{\mathrm{pl}}}
\end{eqnarray}

From these expressions, one can derive the mass $M$ required for each
star of the hidden binary system to lead to the observed values for a
pseudo-planet:

\begin{equation} \label{eq:mass_supposed}
M =  (4.5)^{-3/5} \, M^{3/5}_{\mathrm{pl}} \, M^{2/5}_{*} \, 
\left( \frac{a_{A}}{a_{pl}} \right)^{12/5}
\end{equation}

\section{Application to exoplanets detected by radial velocity}

One may wonder if the low amplitude wobble detected more than 150 stars
(for a permanent update, see {\rm http://www.obspm.fr/planets})
is due to a planet or to a more distant binary system. From the point
of view of radial velocity measurements, a star is considered as
single if there is no long term drift in its velocity curve. 

In our model, we take a star suppossed to be single with a planet
companion but in fact it is a star in a triple system. The absence of
velocity drift imposes a minimum value for the distance  of an
hypothetical companion. $\gamma = GM/a^2_{A}$ being the acceleration
of the target star due to a companion at a distance $a_{A}$, 
the velocity drift acquired over a time span $\Delta T$ is
$\Delta V = \gamma \Delta T = GM/a^2_{A} \Delta T$. The star is single
if $\Delta V$ is smaller than the the observational limit. Taking from
the last years of radial velocity surveys  $\Delta V < 10$ m/s and
$\Delta T$ = 5 yrs, on gets, for $M=1M_{\odot}$, $a_{A} > 300$ pc.

From equations \ref{wobble}, \ref{eq:period}, \ref{eq:sem_maj_ax_pl}
and \ref{eq:mass} we obtain a relation between the mass of the
supposed planet and the period of the wobble (with $M = M_{*} = 1$
solar mass):

\begin{equation} \label{eq:mass_planet}
M_{\mathrm{planet}} \left( M_{\oplus} \right) = \left( \frac{35 \,
  AU}{a_{A}} \right)^{4} \left( \frac{P_{*}}{1 \, \mathrm{year}}
  \right)^{8/3}
\end{equation}

\begin{figure}[th]
  \begin{center}
  \includegraphics[width=0.8\linewidth]{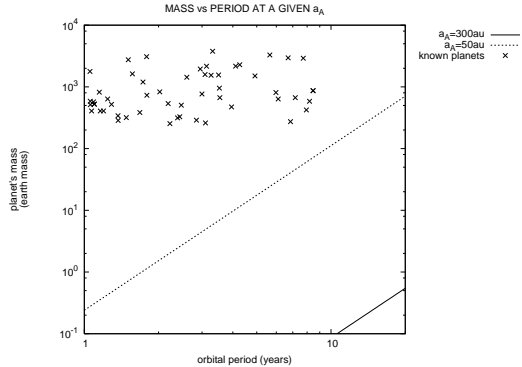}
  \end{center}
  \caption{The perturbation caused by the binary system in the target
  star will imitate a planet whose mass (for a given orbital period)
  lays in the straight lines, which represent distances from the
  target star to the center of mass of the binary system of $50$ and
  $300$ AU.}
  \label{fig:massvsperiod}
\end{figure}

In fig.\ref{fig:massvsperiod} we represent this relation for $a_{A} =
300$ AU and $a_{A} = 50$ AU together with the data for most of the
known extrasolar planets known nowadays. Clearly, none of the planets
found risk of being a triple system, because the effect that a binary
system causes is small. 

\subsection{Planets in binary star systems}

There are presently around 15 planets  detected in binary 
star systems (Eggenberger et al. \cite{eggenberger04}). One may wonder
to what extent the companion to the target star hosting a planet is in
fact a binary system inducing a stellar wobble imitating the effect of
a planet.

Since the separation $a_A$, the orbital parameters of the planet and
the mass of the star (given its spectral type) are then known; the
mass of the hypothetical binary is given by equation
\ref{eq:mass_supposed}. In other words, let's suppose that there is no
planet and that in fact there is a binary system perturbing the motion of
the target star. We can calculate the mass of this system to see if
this hypothesis is reliable. From the data of Eggenberger et
al. \cite{eggenberger04}, table \ref{tb:planets} shows in each case
the value for $M$ derived from equation \ref{eq:mass_supposed}.

\begin{table}
\begin{minipage}[t]{\columnwidth}
\caption[]{Experimental data for planets in binary systems. Last
  column is the mass (calculated with \ref{eq:mass_supposed}) for each
  star of an hypothetical binary system which would cause in the
  target star same wobble as the planet (see text). Data taken from
  Eggenberger et al. \cite{eggenberger04} and from the Extrasolar
  Planets Encyclopaedia ({\rm http://www.obspm.fr/planets}).}
\label{tb:planets}
\centering
\renewcommand{\footnoterule}{}
\begin{tabular}{lcccc}
\hline\hline
{\bf Name} & 
  $a_A \mathrm{(AU)}$\footnote{semimajor axis of the orbit of the
    binary stellar system in AU.} &
  $a_{pl} \mathrm{(AU)}$\footnote{semimajor axis of the orbit of the
    planet in AU.} &
  $M_{pl}\,(M_{J})$\footnote{mass of the planet in Jupiter masses.} & 
  $M\,(M_{\sun})$\footnote{mass calculated with \ref{eq:mass_supposed}
    for each star of the hypothetical binary system which would
    imitate the wobble of a planet.} \\ 
\hline
HD 40979       & $6\,400$ & 0.811 &  3.32  & $3.0\cdot10^{7}$ \\
GL 777 A       & $3\,000$ & 4.8   &  1.33  & $3.6\cdot10^{4}$ \\
HD 80606       & $1\,200$ & 0.469 &  3.90  & $2.0\cdot10^{6}$ \\
55 Cnc         & $1\,065$ & 0.115 &  0.84  & $1.8\cdot10^{7}$ \\
               &          & 0.24  &  0.21  & $1.4\cdot10^{6}$ \\
               &          & 5.9   &  4.05  & $3.7\cdot10^{3}$ \\
16 Cyg B       &      850 & 1.6   &  1.5   & $2.8\cdot10^{4}$ \\
$\upsilon$ And &      750 & 0.83  &  2.11  & $1.3\cdot10^{5}$ \\
               &          & 2.50  &  4.61  & $1.5\cdot10^{4}$ \\
HD 178911 B    &      640 & 0.32  &  6.292 & $1.6\cdot10^{6}$ \\
$\tau$ Boo     &      240 & 0.05  &  4.08  & $1.1\cdot10^{7}$ \\
HD 195019      &      150 & 0.14  &  3.51  & $2.5\cdot10^{5}$ \\  
HD 114762      &      130 & 0.35  & 11.03  & $3.9\cdot10^{4}$ \\
HD 19994       &      100 & 1.33  &  1.78  & $2.1\cdot10^{2}$ \\
$\gamma$ Ceph  &       22 & 2.03  &  1.59  & $1.9\cdot10^{0}$ \\
Gl 86          &       20 & 0.11  &  4.0   & $3.5\cdot10^{3}$ \\
\hline
\end{tabular}
\end{minipage}
\end{table}

The values found for $M$ are aberrant (except in the case of $\gamma$
Ceph); thus the corresponding  stellar wobble cannot be explained as
due to the binary nature of the companion star. In $\gamma$ Ceph the
value of 2 solar masses is not aberrant, however, the mass found for
the companion is 0.4 solar masses (Dvorak et al. \cite{dvorak03}). Our
model is consistent with a binary system of two solar masses each
orbiting at 2.4 AU. That system has not been discovered, observational
results do not point in that direction; we discourage the existence of
the triple system and accept the presence of a planet.

\section{Application to future astrometric searches}

One may wonder if in the
future this problem can become serious with future more accurate measurements
of stellar wobbles. Clearly, from fig. \ref{fig:massvsperiod}, the wobble
indiced by a distant binary star can merely mimic low mass planets on wide 
orbits. This configuration escapes planet detection by radial velocity but
is well adapted to astrometric detection.

\begin{figure}[ht]
  \begin{center}
  \includegraphics[width=0.8\linewidth]{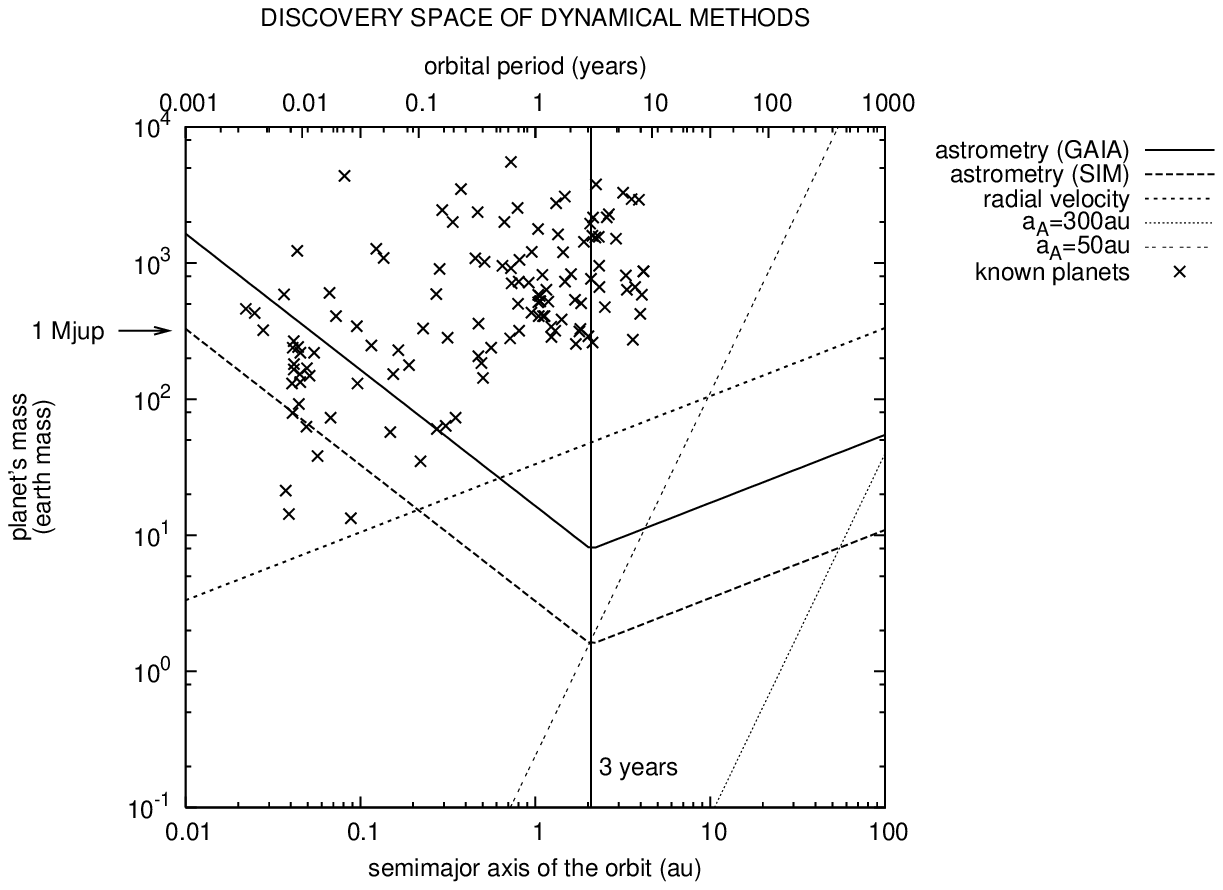}
  \end{center}
  \caption{Discovery space for GAIA, SIM and radial velocity missions,
  together with the expected perturbation caused by binary systems at
  50 and 300 AU.}  
  \label{fig:future2}
\end{figure}

In fig.\ref{fig:future2} we represent the expected discovery space for
two spatial missions: GAIA (Sozzetti et al. \cite{sozzetti03}) and
SIM (Space Interferometry Mission, JPL). The constraints are the limiting
resolution of 10 and 2 micro arcseconds respectively and the
lifetime of the missions: three years. Both will detect planets in
regions \textbf{still} unexplored by radial velocity methods. 

\begin{figure}[ht]
  \begin{center}
  \includegraphics[width=0.8\linewidth]{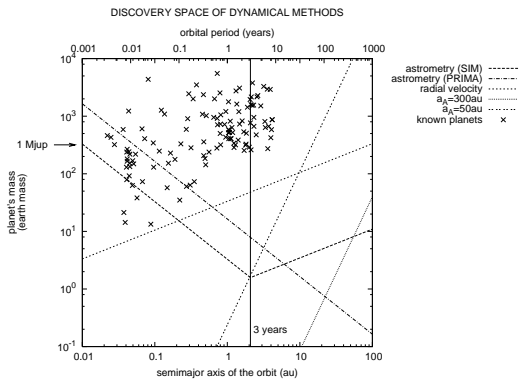}
  \end{center}
  \caption{Discovery space for SIM, PRIMA and radial velocity missions,
  together with the expected perturbation caused by binary systems at
  50 and 300 AU.} 
  \label{fig:future3}
\end{figure}

In fig.\ref{fig:future3} we compare the expected results for SIM and
PRIMA (Phase-Referenced Imaging and Microarcsecond Astrometry at ESO
VLTI). PRIMA has the same resolution as GAIA (10 micro arcseconds) but
is not constrained by the three years lifetime, having access to a
low-mass long-period region where this stellar wobble effect will be
more important.

The possibility of stellar wobble simulating a planet is small,
however, for long periods and if the distance from the target star to
the binary system ($a_{A}$) is not that important (e.g. 50 AU), those
effects will have to be taken into account.

\section{Conclusion}

By itself a stellar wobble is not  a proof that a planet is 
detected. It is necessary to verify that no far binary star generates the
wobble or to confirm that planet by transit or direct imaging
observations. For the presently know planets, the explanation by a
perturbing binary star can nevertheless be ruled out. But the
sensitivity of GAIA, PRIMA and SIM is such that for some regions of the
($M_{pl}$, $P_{pl}$) plane there can be an ambiguity between a true
planet detected by astrometry and a wobble induced by a binary star.

%
\begin{acknowledgements}

We are grateful to R. Dvorak for the permission of the use of his
3-body numerical simulation software KAPPA.

This research has made use of the SIMBAD database, operated at CDS,
Strasbourg, France.
\end{acknowledgements}

%
\clearpage

\onecolumn

\appendix

\section{}
The perturbation of the trajectory of a body in a triple system by a
distant binary system is a classical issue in the 3-body problem
(Roy \cite{roy}). Here we remind the analysis only for
self-consistency of our paper.

Consider a simplified triple hierachical system consisting of a binary
system of two equal mass bodies ($M_{1} = M_{2} = M$) in a circular
orbit with radius $a_{B}$ plus a third companion orbiting the center
of mass of the binary system in a bigger circular orbit with radius 
$a_{A} \gg a_{B}$.

Suppose that the motion of the binary system is not perturbed by
the third body (let $M_{3}=0$) and let them move in the plane
$z=0$. As we want a circular orbit with a given angular velocity
$\omega$, we find that the equation of motion of the bodies $M_{1}$
and $M_{2}$ is:
\begin{eqnarray} \nonumber 
\overrightarrow{r}_{M_{1}} & = & a_{B} \left( \cos{\omega t} \, \hat{i} + 
  \sin{\omega t} \, \hat{j} \right)
\\ \label{eq:def_rMi}
\overrightarrow{r}_{M_{2}} & = & - a_{B} \left( \cos{\omega t} \, \hat{i}
  + \sin{\omega t} \, \hat{j} \right)
\\ \nonumber
\omega & = & \sqrt{\frac{2 G M}{(2 a_{B})^{3}}} = 
  \sqrt{\frac{G M}{4 a_{B}^{3}}}
\end{eqnarray}

The gravitational force per unit of mass in any point
$\overrightarrow{r}=(x,y,z)$ caused by these two bodies of \textbf{mass} $M$
is:
\begin{equation} \label{eq:newton} 
\overrightarrow{F} = - G M \left( 
  \frac{\overrightarrow{r_{1}}}{|\overrightarrow{r_{1}}|^{3}} + 
  \frac{\overrightarrow{r_{2}}}{|\overrightarrow{r_{2}}|^{3}} \right)
  \qquad \textrm{where} \,\,
  \overrightarrow{r}_{i} = \overrightarrow{r} -
  \overrightarrow{r}_{M_{i}} \quad i=1,2
\end{equation}

To make the problem even simpler, we are going to suppose that the
motion of the third star is in the plane defined by the orbit of the
binary system. In the plane we keep only the cartesian coordinates
$(x,y)$ or their polar equivalent $(r,\theta)$ with the identities for
the unitary vectors:
\begin{eqnarray} \nonumber 
\hat{r} & = & \cos \theta \, \hat{i} + \sin \theta \, \hat{j}
\\ \nonumber
\hat{\theta} & = & - \sin \theta \, \hat{i} + \cos \theta \, \hat{j}
\end{eqnarray}

then we can separate \ref{eq:newton} into the central force (in the
direction of the center of mass of the binary system, parallel to
$\hat{r}$) and the angular force (perpendicular to the former,
parallel to $\hat{\theta}$):
\begin{eqnarray} \nonumber 
\overrightarrow{r}_{1/2} = \overrightarrow{r} -
  \overrightarrow{r}_{M_{1/2}} & = & \left[ r \cos{\theta} \, \hat{i}
  + r \sin{\theta} \, \hat{j} \right] - \left[ \pm a_{B} \left(
  \cos{\omega t} \, \hat{i} + \sin{\omega t} \, \hat{j} \right)
  \right] = 
\\ \nonumber
& = & \left[ r \mp a_{B} \cos{ \left( \omega t - \theta \right) }
  \right] \hat{r} \mp a_{B} \sin{ \left( \omega t - \theta \right) }
  \hat{\theta}
\end{eqnarray}

\begin{displaymath} 
| \overrightarrow{r}_{1/2} |^{2} = r^{2} + a_{B}^{2} \mp 2 r a_{B}
  \cos{ \left( \omega t - \theta \right) }
\end{displaymath}

\begin{displaymath} 
\frac{1}{ | \overrightarrow{r}_{1/2}^{3} | } \approx
  \frac{1}{r^{3}} \left[ 1 \pm 3 \, \epsilon \cos{ \left( \omega t -
  \theta \right) } + \frac{15}{2} \, \epsilon^{2} \cos^{2}{ \left(
  \omega t - \theta \right) } - \frac{3}{2} \, \epsilon^{2} \right]
\qquad
\mathrm{where} \,\, \epsilon \equiv  \frac{a_{B}}{r} \ll 1
\end{displaymath}

\begin{eqnarray} \nonumber 
\overrightarrow{F} = \ddot{\overrightarrow{r}} & = & - G M \left( 
\frac{\overrightarrow{r_{1}}}{|r_{1}|^{3}} + 
\frac{\overrightarrow{r_{2}}}{|r_{2}|^{3}} \right) \approx
\\ \label{eq:newton2}
& \approx & - \frac{2 G M}{r^{2}} \left( \left[ 1 + \frac{3}{4} \,
  \epsilon^{2} + \frac{9}{4} \, \epsilon^{2} \cos 2 (\omega t -
  \theta) \right] \hat{r} - \frac{3}{2} \, \epsilon^{2} \sin 2
  (\omega t - \theta) \hat{\theta} \right)
\end{eqnarray}

correct to order 2 in $\epsilon$ and which is the same expression we
can find in equation $14.25$ of (Roy \cite{roy}) (under our assumption
that the masses of the binary system are equal and the third mass is
zero). As $\ddot{\overrightarrow{r}} = \left( \ddot{r}-r
\dot{\theta}^{2} \right) \hat{r} + \left( 2 \dot{r} \dot{\theta} + r
\ddot{\theta} \right) \hat{\theta}; $ we finally arrive to:
\begin{eqnarray} \nonumber 
\ddot{r}-r \dot{\theta}^{2} & \approx &
  - \frac{2 G M}{r^{2}} 
  \left[ 1 + \frac{3}{4} \epsilon^{2} + \frac{9}{4} \epsilon^{2} \cos
  2 (\omega t - \theta) \right]
\\ \label{eq:systeme}
2 \dot{r} \dot{\theta} + r \ddot{\theta} & \approx &
  \frac{3 G M}{r^{2}} \epsilon^{2} \sin{ 2 \left( \omega t - \theta
  \right) }
\end{eqnarray}

As a solution to the system of differential equations
(\ref{eq:systeme}) we propose that the third body is making a circular
orbit of radius $a_{A}$ around the center of mass of the binary system
with angular velocity $\Omega$ perturbed by another elliptical motion
of semiaxes $\delta_{x}$ and $\delta_{y}$ both much smaller than
$a_{A}$. So the equation of motion of this body is, in rectangular
coordinates:
\begin{eqnarray} \nonumber
x(t) & = & a_{A} \cos{\Omega t} + \delta_{x} \cos{2 \omega t}
\\ \label{eq:pos_M3}
y(t) & = & a_{A} \sin{\Omega t} - \delta_{y} \sin{2 \omega t}
\end{eqnarray}

Notice:
\begin{enumerate}
  \item this is not the most general solution. In fact, it is closer
  to a perturbation solution: these equations are solution of the
  differential equation system (\ref{eq:systeme}).
\item the angular velocity of the perturbation is $-2 \omega$. As the
  two bodies of the binary system have the same mass, every half a
  revolution of the system the third body \emph{sees} the same
  configuration of the binary system, in other words, the same
  configuration of the perturbation.
\item the Keplerian angular velocity goes as $a^{-1.5}$. As $a_{A} \gg
  a_{B} \Rightarrow \omega \gg \Omega$. 
\end{enumerate}

Taking this into account we arrive to the following equations to order
one in the perturbation (that is, taking into account that
$\delta_{x}/a_{A} \ll 1; \; \delta_{y}/a_{A} \ll 1; \; \Omega/\omega
\ll 1;$):
\begin{equation} \label{eq:solution_systeme}
\begin{array}{ll}
  r \equiv \sqrt{ x^{2} + y^{2} } \approx a_{A} + \delta_{x} \cos{2
  \omega t}
  &
  \qquad \theta \equiv \arctan{\frac{y}{x}} \approx \Omega t
\\
  \dot{r} \equiv \frac{d}{dt} \sqrt{ x^{2} + y^{2} } \approx
  - 2 \omega \delta_{x} \sin{2 \omega t}
  &
  \qquad \dot{\theta} \equiv \frac{x \dot{y} - \dot{x} y}{r^{2}}
  \approx \Omega - 2 \omega \frac{\delta_{y}}{a_{A}} \cos{2 \omega t}
\\ 
  \ddot{r} \equiv \frac{d^{2}}{dt^{2}} \sqrt{ x^{2} + y^{2} } \approx
  - 4 \omega^{2}  \delta_{x} \cos{2 \omega t}
  &
  \qquad \ddot{\theta} \equiv \frac{d}{dt} 
  \frac{x \dot{y} - \dot{x} y}{r^{2}} \approx 4 \omega^{2}
  \frac{\delta_{y}}{a_{A}} \sin{2 \omega t}
\end{array}
\end{equation}

Introducing this solution (\ref{eq:solution_systeme}) in the system of
equations (\ref{eq:systeme}) we obtain the following identities:
\begin{eqnarray} \nonumber
a_{A} \Omega^{2} + 4 \omega^{2} \delta_{x} \cos{2 \omega t} & = & 
  \frac{2 G M}{a_{A}^{2}} \left(  1 + \frac{3}{4} \epsilon^{2} +
  \frac{9}{4} \epsilon^{2} \cos{2 \omega t} \right)
\\ \nonumber
4 \omega^{2} \delta_{y} \sin{2 \omega t} & = & 
  \frac{3 G M}{a_{A}^{2}} \epsilon^{2} \sin{2 \omega t}
\end{eqnarray}

from wich we arrive to:

\begin{equation} \label{wobble}
  \delta_{x} = 4.5 \,  \frac{a_{B}^{5}}{a_{A}^{4}}
\qquad \qquad
  \delta_{y} = 3  \, \frac{a_{B}^{5}}{a_{A}^{4}}
\qquad \qquad
  \Omega^{2} = \frac{2 G M}{a_{A}^{3}} \left( 1 + 0.75
  \frac{a_{B}^{2}}{a_{A}^{2}} \right)
\end{equation}

The mass of the binary system does not affect the
amplitude of the perturbation but does affect the period
($T_{\textrm{\small{wobble}}} \sim (2 \omega)^{-1} \sim
M^{-0.5}$). The angular velocity $\Omega$ is bigger than the Keplerian
one, that means that for the same mass, the system is less bounded
(the total energy $E=T-U$ is \emph{less negative}).

%


\begin{thebibliography}{}

\bibitem[2003]{dvorak03}
Dvorak, R., Pilat-Lohinger, E., Funk, B., Freistetter, F.,
2003. Planets in habitable zones: a study of the binary Gamma Cephei.
A\&A 398, L1-L4.

\bibitem[2004]{eggenberger04}
Eggenberger A., Udry S. \& Mayor M., 2004. Statistical properties of
exoplanets. III. Planet properties and stellar multiplicity. A\&A 417,
353-360. 

\bibitem[1979]{roy} 
Roy, A. E., 1979, Orbital Motion. Adam Hilger Ltd, Bristol.

\bibitem[1999]{schneider99}
Schneider, J., 1999, The wobble method of extrasolar planets detection
revisited. American Astronomical Society, DPS meeting 31, No. 4, \#
5.02

\bibitem[2003]{sozzetti03}
Sozzetti A., Casertano S.,  Lattanzi M. \& Spagna A., 2003. The GAIA
astrometric survey of the solar neighborhood and its contribution to
the target database for DARWIN/TPF. In {\em Proceeedings of the
conference on Toward Other Earths: DARWIN/TPF and the Search for
Extrasolar Terrestrial Planets}. ESA SP-539, Noordwijk, Netherlands,
p. 605-610.

\end{thebibliography}
\end{document}